\documentclass[preprint,twocolumn]{aastex}
\usepackage{epsf}
\usepackage{graphicx}
\usepackage{emulateapj5}
\usepackage{onecolfloat}
\usepackage{apjfonts}
\usepackage{amsmath}
\usepackage{aalongtable}

\newcommand{\lsim}{\lower0.6ex\vbox{\hbox{$ \buildrel{\textstyle <}\over{\sim}\ $}}}
\newcommand{\gsim}{\lower0.6ex\vbox{\hbox{$ \buildrel{\textstyle >}\over{\sim}\ $}}}

\newcommand{\hkpc}{\,{h^{-1}\mathrm{kpc}}}

\newcommand{\hMsun}{\,{h^{-1}\mathrm{M}_{\odot}}}
\newcommand{\hMpc}{\,{h^{-1}\mathrm{Mpc}}}
\newcommand{\Mpc}{\,{\mathrm{Mpc}}}
\newcommand{\kpc}{\,{\mathrm{kpc}}}

\newcommand{\kms}{{\,{\mathrm{km}}\,{\mathrm{s}}^{-1}}}

\newcommand{\mpt}{m_{\mathrm{p}}}
\newcommand{\seconds}{\,{\mathrm{s}}}

\newcommand{\Omegam}{\Omega_{M}}
\newcommand{\Omegab}{\Omega_{B}}
\newcommand{\Omegal}{\Omega_{\Lambda}}

\newcommand{\sig}{\sigma_{8}}
\newcommand{\rhomean}{\rho_{\mathrm{M}}}

\newcommand{\Rvir}{R_{180}}
\newcommand{\Mvir}{M_{180}}

\newcommand{\Vcirc}{V_{\mathrm{circ}}}
\newcommand{\Vmax}{V_{\mathrm{max}}}

\newcommand{\rfind}{r_{\mathrm{find}}}
\newcommand{\rt}{r_{\mathrm{t}}}
\newcommand{\rp}{r_{\mathrm{p}}}

\newcommand{\bmag}{M_{\mathrm{B}}}

\newcommand{\Pks}{P_{\mathrm{KS}}}

\newcommand{\Nsats}{N_{\mathrm{sats}}}

\newcommand{\Nprog}{N_{\mathrm{prog}}}
\newcommand{\fprog}{f_{\mathrm{prog}}}

\newcommand{\Nretr}{N_{\mathrm{retr}}}

\newcommand{\ffalse}{f_{\mathrm{false}}}

\newcommand{\rtrue}{r_{\mathrm{true}}}

\newcommand{\beq}{\begin{equation}}
\newcommand{\eeq}{\end{equation}}

\newcommand{\dd}{\mathrm{d}}

\begin{document}
\submitted{The Astrophysical Journal, submitted}
\vspace{1mm}
\slugcomment{{\em The Astrophysical Journal, submitted}} 
\shortauthors{Azzaro et al.}
\twocolumn[

\lefthead{Satellites of Disk Galaxies}
\righthead{Azzaro, Zentner, Prada, \& Klypin}
\title{The peculiar velocities of satellites of external disk galaxies}
\author{Marco Azzaro\altaffilmark{1,2}, 
Andrew R. Zentner\altaffilmark{3}, 
Francisco Prada\altaffilmark{2,4,5},
Anatoly A. Klypin\altaffilmark{6}
}
%
%
\begin{abstract}

We analyze the angular distribution and the orbital rotation
directions of a sample of carefully-selected satellite galaxies
extracted from the Sloan Digital Sky Survey (SDSS).  We also study
these statistics in an $N$-body simulation of cosmological structure
formation set within the $\Lambda$CDM paradigm under various
assumptions for the orientations of disk angular momenta.  Under the
assumption that the angular momenta of the disks are aligned with the
angular momenta of the inner regions of their host dark matter halos,
we find that the fraction of simulated satellite halos that exhibit
prograde motion is $\fprog \approx 0.55-0.60$, with larger satellites
more likely to be prograde.  In our observational sample,
approximately $60\%$ of the satellites exhibit prograde motion, a
result that is broadly consistent with the simulated sample.  Contrary
to several recent studies, our observational sample of satellite
galaxies show no evidence for being anisotropically distributed about
their primary disks.  Again, this result is broadly consistent with
our simulated sample of satellites under the assumption that disk and
halo angular momenta are aligned. However, the small size of our
observational sample does not yet allow us to distinguish between
various assumptions regarding the orientations of disks in their
halos.  Finally, we assessed the importance of contamination by
interlopers on the measured prograde and retrograde statistics.
\end{abstract}
]

%
%
%
%

\altaffiltext{1}{
Universidad de Granada, Granada, Spain
}
\altaffiltext{2}{
Isaac Newton Group of Telescopes, Santa Cruz de La Palma, Spain
}
\altaffiltext{3}{
Kavli Institute for Cosmological Physics and 
Department of Astronomy and Astrophysics, 
The University of Chicago, Chicago, IL, USA
}
\altaffiltext{4}{
Instituto de Astrof{\'{\i}}sica de Canarias, La Laguna, Tenerife, Spain
}
\altaffiltext{5}{
Ram{\'{o}}n y Cajal Fellow, 
Instituto de Astrof{\'{\i}}sica de Andaluc{\'{\i}}a, 
Granada, Spain
} 
\altaffiltext{6}{
Astronomy Department, 
New Mexico State University, 
Las Cruces, NM, USA
}

%
%
%
%
%
%
%
%
\section{Introduction}
\label{sec:intro}

In the standard hierarchical, 
cold dark matter (CDM) scenario of 
cosmological structure and galaxy formation 
\citep[e.g.][]{white_rees78,blumenthal_etal84}, 
large systems are generally formed 
through the continual merging of comparably early-forming, 
low-mass objects in the hierarchy.  
In this context, the halos of large galaxies are 
formed by smaller halos that are accreted by the 
larger host and disrupted by tidal interactions.  
As the massive galactic halo is assembled, 
other halos and their associated 
galaxies may become gravitationally 
bound to the host and orbit in the host potential before 
possibly being incorporated into the central host galaxy 
or being strongly affected by tides 
\citep[e.g.][]
{taffoni_etal03,hayashi_etal03,zentner_bullock03,
kravtsov_etal04,kazantzidis_etal04b,
taylor_babul04,zentner_etal05}.  

The distribution of satellite galaxies about their 
primary galaxies has received significant attention, 
both theoretically and observational.  A number of 
studies have focused on the radial distributions of 
satellite halos and satellite galaxies 
\citep[e.g.][]
{taylor_etal03,zentner_bullock03,kravtsov_etal04,
diemand_etal04,gao_etal04a,vdb_etal05,
zentner_etal05,nagai_kravtsov05}; however, the 
angular distribution of satellites has become a 
subject of a great deal of recent activity.
In an early study, Holmberg \citep{holmberg69}, 
found that satellites of spiral 
galaxies with projected separations 
$\rp \lsim 50 \kpc$ are preferentially located near 
the short axes of the projected light distributions 
of their host galaxies.  In other words, they are 
preferentially located near the poles of the primary 
disk galaxies.  \citet{zaritsky_etal97b} found a 
statistically-significant anisotropy, similar to that 
advocated by \citet{holmberg69}, at larger projected 
separations ($200 \kpc \lsim \rp \lsim 500 \kpc$) and, 
in a more recent study, \citet{sales_lambas04} found 
evidence for the preferential alignment of satellites 
along the minor axes of their primaries in the 
Two Degree Field Galaxy Redshift Survey 
\citep[e.g.][]{colless_etal01}.  However, the observational 
status of the so-called {\em Holmberg Effect} is unclear.  
\citet{brainerd04} studied satellites in the 
Sloan Digital Sky Survey 
\citep[SDSS, e.g.][]{york_etal00,strauss_etal02} and 
found evidence for the {\em opposite} correlation of satellite 
position with the {\em major} axis of the light distribution 
of the primary galaxy.  

\citet{kroupa_etal05} recently reiterated that 
the satellite galaxies of the Milky Way (MW) 
are distributed anisotropically 
\citep[see also][]{lynden-bell82,Majewski94,Hartwick96,
Hartwick00,mateo98,Grebel_etal99,Willman_etal04} and 
argued that this observation presents a challenge for 
the standard CDM paradigm.  
\citet[][see also \citeauthor{kang_etal05} \citeyear{kang_etal05}; 
\citeauthor{libeskind_etal05} \citeyear{libeskind_etal05};
\citeauthor{agustsson05} \citeyear{agustsson05};]{zentner_etal05b} 
showed that MW-sized dark matter halos contain satellite 
halos or {\em subhalos} that are distributed anisotropically 
about their central {\em host} halos.  Subhalos are 
located preferentially near the long axes of their host dark 
matter halos, so that the anisotropy of MW satellites can be 
understood if the rotation axis of the MW disk is closely 
aligned with the long axis of the triaxial host dark matter 
halo of the MW.  In order to test this alignment 
{\em conjecture}, it is necessary to examine a 
large sample of satellites about disk galaxy primaries.

Another interesting way to relate the properties of satellite 
galaxies to disk primaries is through the sense of 
rotation of the satellites relative to the disks, if 
any.  On the theoretical side, the prediction for the 
number of prograde and retrograde satellites is unclear 
and there are many aspects to consider.  
In the simplest picture of disk galaxy formation, 
baryons in halos start off sharing the angular momentum 
distribution of their halos and, on average, conserve 
angular momentum as they cool and condense
\citep[e.g.][]{fall_efstathiou80}.  This 
implies that the poles of disk galaxies should be 
collinear with the net angular momentum vectors of 
their host halos which, in turn, are generally aligned 
with the halo minor axes 
\citep[e.g.][]{warren_etal92,porciani_etal02,
faltenbacher_etal05}.  
\citet[][see also \citeauthor{maller_etal02} 
\citeyear{maller_etal02}]{vitvitska_etal02} 
proposed that the angular momenta 
of dark matter halos are built through mergers with 
infalling satellite halos.  
If the accretion of satellites is a random process and 
the acquisition of angular momentum is a random walk as in the 
\citet{vitvitska_etal02} model, this scenario seems to favor 
neither net prograde nor net retrograde satellite rotation 
because the angular momentum of the halo at disk formation may 
bear little resemblance to the angular momenta carried by the 
late-accreting satellites.  However, the accretion of satellite 
halos generally occurs along a preferred direction, so that 
the accretion directions of satellites are strongly correlated 
\citep[e.g.][]{knebe_etal04,zentner_etal05b}, the detailed 
consequences of which are unclear.  

There are also dynamical processes that may affect the prograde 
and retrograde fractions of satellites subsequent to satellite 
accretion.  \citet{quinn_goodman86} explored the idea that 
enhanced dynamical friction for satellites on 
nearly coplanar, prograde orbits about the disk could lead to 
enhanced destruction of prograde satellites but found this 
process to be fairly inefficient.  
\citet{penarrubia_etal02} extended this to include the effect of 
an oblate halo about the disk; however, 
their result was that preferred destruction 
is inefficient for satellites further than $\sim 50 \kpc$ 
from the primary because the orbital decay times are large 
at these radii \citep[see also][]{zaritsky_gonzalez99}.  
Moreover, CDM halos generally tend 
to be more prolate rather than oblate 
\citep[e.g.][]{jing_suto02,bullock02,kazantzidis_etal04}.
Another dynamical argument relates to the heating of galactic 
disks.  \citet{velazquez_white99} conducted 
self-consistent $N$-body simulations of satellites 
interacting with disk galaxies and found that massive 
satellites may be accreted without destroying the primary 
disk galaxy, 
particularly for satellites on retrograde orbits which 
produce significantly less disk heating than satellites 
on prograde orbits.  

>From an observational point of view, 
\citet{carignan_etal97} found that seven satellites 
around the giant lenticular galaxy NGC5084 exhibit 
retrograde motion while only one satellite is prograde.  
Moreover, the large mass and the tilted disk of NGC5084 
suggest that this galaxy has survived a number of mergers, 
possibly with prograde satellites.  Similarly, 
\citet{zaritsky_etal97} examined the orbits of their 
sample of satellites about disk galaxies and found that 
a slight excess ($52\%$) of satellites exhibit 
retrograde motions in a sample of 
$69$ primaries with $115$ satellites.

Any statistical work on satellite galaxies has to face the problem of
the small average number of satellites that can be identified for each
primary \citep[e.g.][]{zaritsky_etal97,mckay_etal02,prada_etal03}.
Provided that the primary galaxies are selected in a consistent and
homogeneous way, it is useful to consider all satellites in the sample
as satellites of a single, fictitious primary.  This is the assumption
that we work under here.

In this paper we use a sample of hosts and satellites extracted from
the SDSS \citep{prada_etal03} and a sample of halos and subhalos in a
cosmological $N$-body simulation in order to address a number of
questions concerning the angular momenta of disk galaxies relative to
their satellites as follows.
\begin{enumerate}
\item What is the distribution of angular distances 
of satellites from the planes of the primary disk 
galaxies and how does this compare to the 
theoretically-predicted angular distribution of 
satellite halos in simulations of structure formation 
in the context of CDM?

\item What is the ratio between prograde and 
retrograde satellites ($P/R$) in a well-defined sample 
of observed satellite systems and how does this 
compare to predicted satellite orbits? 

\item Does the fraction of prograde satellites 
depend on the angular distance of the satellites 
from the plane of the primary disk or with 
projected physical separation?  

\item How does the presence of interlopers 
(objects that are field galaxies, 
not dynamically bound to the primary, 
but counted as satellites due to projection) 
affect $P/R$ in our observational sample 
relative to the value for ``true'' satellites?  

\end{enumerate}
This manuscript is organized as follows.  
In the next section, 
we describe the numerical simulations and 
analysis methods that we use to compute theoretical 
predictions for satellite distributions and present 
the basic results from the simulation.  
We describe our data and present the observational
results in \S~\ref{sec:observations}.
In \S~\ref{sec:biases}, we discuss the
effects of interlopers and observational biases.  
Finally, we present our results and draw our conclusions 
in \S~\ref{sec:results}. The main sample from which our
objects are extracted \citep[see][]{prada_etal03} 
was compiled using $B$ magnitudes,
therefore, throughout this paper, we refer exclusively
to $B$ magnitudes, either apparent or absolute.
%
%
%
%
\section{Theoretical Predictions}
\label{sec:predictions}

%
%
\subsection{Numerical Simulation}
\label{sub:sims}

To compare the observed satellite sample to the 
theoretically-predicted distributions of satellites 
in the context of CDM, we analyzed a simulation of 
structure formation in the so-called concordance 
cosmological constant plus CDM 
cosmology ($\Lambda$CDM), with $\Omegam=0.3$, 
$\Omegal=0.7$, $h=0.7$, $\Omegab h^2 = 0.22$, 
$n_{\mathrm{s}}=1$, and a 
power spectrum normalization of $\sig = 0.9$. 
The simulation followed the evolution of 
$512^3$ particles in a computational box of 
$80 \hMpc \simeq~114.3 \Mpc$ on a side, 
implying a particle mass of 
$\mpt \simeq~3.16 \times 10^8 \hMsun$.  

The simulation was performed using the Adaptive 
Refinement Tree $N$-body simulation code 
\citep[ART][]{kravtsov_etal97,kravtsov99}.  
The ART code employs a particle-mesh approach 
and achieves high force resolution 
by refining the grid in all high-density 
regions with a recursive refinement algorithm.  
This creates an hierarchy of refinement 
meshes of different resolution covering 
regions of interest.  The mesh cells were 
refined up to a maximum of eight levels, 
corresponding to a minimum cell size 
of $h_{\mathrm{peak}} \simeq~1.22 \hkpc$.

We identified halos and subhalos using a variant 
of the Bound Density Maxima algorithm
\citep{klypin_etal99b}.  We began by computing the 
local density at each particle using a smoothing 
kernel of $24$ particles and identified local maxima 
in the density field.  We proceeded through all 
density peaks, beginning with the highest density peak 
and moving toward lower density, and marked each peak as 
a potential halo center.  We surrounded each such peak 
with a sphere of radius $\rfind = 20 \hkpc$ and excluded 
all particles within this sphere from further consideration 
as a potential halo center.  The parameter $\rfind$ is 
determined by the size of the smallest halos that we aim to 
identify robustly.  After identifying potential halo centers, 
we iteratively removed unbound particles.  

We used the remaining bound particles to compute the circular 
velocity profile of the halo $\Vcirc = \sqrt{GM(<r)/r}$, 
and the maximum circular velocity $\Vmax$.  
We defined the virial masses and radii of halos 
corresponding to a fixed, 
spherical overdensity of $\Delta = 180$ times the 
{\em mean density} of the universe $\rhomean$, so that 
$\Mvir \equiv 4\pi (\Delta \rhomean) \Rvir^3/3$
\footnote{Note that other conventions that appear 
in the literature define halo mass with respect to 
a {\em virial overdensity} motivated by spherical 
top-hat collapse 
\citep[e.g.][]{lacey_cole93,eke_etal98}, 
which gives an overdensity $\Delta \simeq 178$ in models 
with $\Omegam=1$ and an overdensity that varies with 
redshift in models with $\Omegam < 1$.  In the $\Lambda$CDM 
model that we study, $\Delta \simeq 340$ at $z=0$ and 
$\Delta \rightarrow 178$ at high redshift 
\citep[for a useful fitting function for $\Delta(z)$ see]
{bryan_norman98}.  
The definition $\Delta = 180$ is convenient due 
to the universality of the mass function according 
to this definition 
\citep[see][for details]{jenkins_etal01}}.  
For {\em satellite halos} or {\em subhalos} that are 
self-bound substructures within the virial radius of a 
larger {\em parent} or {\em host} halo, the outer 
boundary is somewhat ambiguous.  We adopted a 
truncation radius $\rt$, at which the slope of the 
density profile became greater than a critical 
value of $\dd \ln \rho / \dd \ln r = -0.5$.  
This criterion was based on the fact that we do not 
expect density profiles of CDM halos to be shallower 
than this and, empirically, this definition is 
approximately equal to the radius at which the 
background density of the host halo particles is equal 
to the density of the particles bound to the subhalo.
In what follows, we choose to quantify the size of 
halos and subhalos according to the peak circular 
velocity $\Vmax$, because for subhalos this 
quantity is measured more robustly and this 
quantity is not subject to the same ambiguity 
as any particular mass definition.

%
%
\subsection{Analysis Methods}
\label{sub:methods}

\begin{figure}[t]
\epsscale{1.0}
\plotone{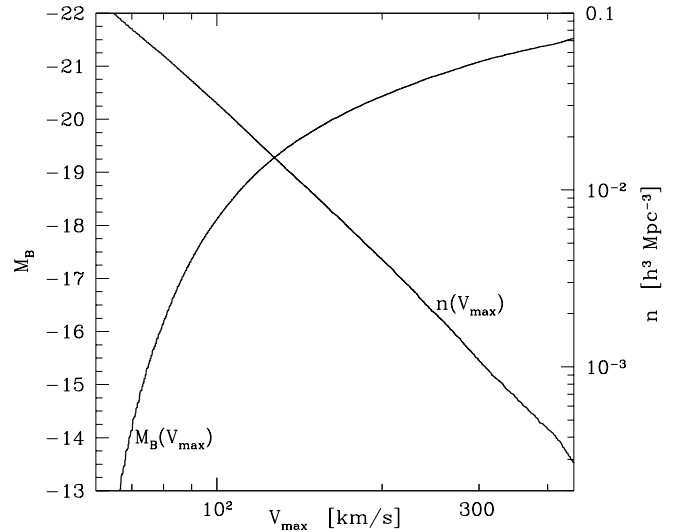}
\caption{
Assigning magnitudes to halos in the $N$-body 
simulation.  The horizontal axis is the maximum 
circular velocity of halos and subhalos $\Vmax$.  
The {\em decreasing function of} $\Vmax$ is the number 
density of halos as a function of maximum 
circular velocity, $n(\Vmax)$.  This line should 
be read against the vertical axis on the {\em right} 
side of the plot.  The {\em increasing function of }
$\Vmax$ is the absolute $B$ magnitude $\bmag$, 
of galaxies with the same number density as the number 
density of halos at this value of $\Vmax$ 
({\em decreasing} line).  This line should be 
read against the vertical axis on the {\em left} 
side of the plot and it represents 
the mapping of halo size ($\Vmax$) to 
absolute magnitude ($\bmag$) that we use 
to compare our observational results 
to halos and subhalos in dissipationless 
simulations.
}
\label{fig:simcat}
\end{figure}

One of our goals is to compare the observed statistics of 
satellite galaxies to theoretical predictions using 
the catalogs of observed satellites and primaries 
described in \S~\ref{sec:observations}.  
In order to do this, we 
construct catalogs of galaxies with absolute $B$ 
magnitudes from the halos and subhalos of the 
simulation.  There are several empirical methods 
to connect galaxies to halos that have been 
explored in the recent literature 
\citep[e.g.][]{seljak00,berlind_weinberg02,yang_etal03,
vdb_etal03a,vdb_etal03b,zheng_etal04,vdb_etal05}.  
For simplicity, we assign magnitudes to halos using 
the prescription advocated by \citet{kravtsov_etal04b} 
in their study of galaxy and halo clustering 
statistics.  We map magnitudes onto halos by 
requiring the number density of halos with maximum 
circular velocities greater than some $\Vmax$ to 
be equal to the observed number density of 
galaxies with absolute $B$ magnitudes greater than 
some corresponding $\bmag(\Vmax)$.  To compute the 
number densities of galaxies of a given magnitude, 
we used the \citet{schechter76} function fits to the 
SDSS $g$-band luminosity function presented in 
\citet{blanton_etal03}.  To convert from 
$g$ to $B$, we used the conversions compiled 
by M. Blanton at 
{\tt URL http://cosmo.nyu.edu/blanton/kcorrect/}.
The resulting correspondence 
between halo $\Vmax$ and $\bmag$ is shown in 
Figure~\ref{fig:simcat}, where we also show the 
number density of halos as a function of $\Vmax$.  
\citet{kravtsov_etal04b} showed that a similar 
mapping of galaxies onto halos using $r$-band 
luminosities reproduces the clustering statistics 
of galaxies as a function of luminosity as measured 
by the SDSS.

%
%

For each primary halo in the catalogs, we computed 
reference vectors that allow us to orient 
satellites relative to halos in a meaningful way.  
To compute these vectors, we used all 
primary halo particles within $R \le 0.3 \Rvir$ 
in order to mitigate the influence of 
large substructures at large halo-centric distances 
on these vectors and to focus on the properties of 
the inner halo which contains the material that 
collapsed earliest and where any disk galaxy 
would reside.  First, we computed the principal 
axes of the triaxial primary halos using an 
iterative algorithm as described in detail in 
\citet[][see also \citeauthor{dubinski_carlberg91} 
\citeyear{dubinski_carlberg91}; 
\citeauthor{kazantzidis_etal04} 
\citeyear{kazantzidis_etal04}]{zentner_etal05b}.  
We also computed the net angular momentum vector 
of all primary halo particles within $R \le 0.3 \Rvir$.
In the following, we compare various hypotheses for 
the orientations of disks in their primary halos to 
the data regarding the relative positions and 
velocities of satellite galaxies in our sample.  
We computed all statistics by summing over 
three orthogonal projections through the 
simulation volume and we computed statistical 
errors by taking each projection to be statistically 
independent.

\subsection{Theoretical Predictions}
\label{sub:simresults}

After assigning magnitudes to halos and determining 
reference vectors for halos, we constructed catalogs 
of primaries and satellites by observing the 
simulation from three orthogonal projection directions 
using the same criteria as we employed for 
our observational sample (see \S~\ref{sub:sample} below).  
Primaries were restricted to only those halos that 
are assigned magnitudes $-20.5 \le \bmag \le -19.5$.  
Primaries were required to satisfy isolation criteria 
such that any halos with projected separations 
$\rp \le 500 \hkpc$ and line-of-sight velocity 
differences $\Delta V \le 1000 \kms$ with respect 
to the primary are dimmer by at least 
$\Delta \bmag \ge 2$.  Satellites were required to 
have $\Delta \bmag \ge 2$, $\rp \le 350 \hkpc$, and 
$\Delta V \le 500 \kms$.

The first results from the simulations concern the 
angular positions of the observed satellites.  
We have examined three {\em idealized} hypotheses 
regarding the orientation of disks in dark matter 
halos:  (1) that the disk rotation axis is aligned with 
the angular momentum of the halo; (2) that the 
disk is aligned with the major axis of the host 
halo; (3) that the disk is aligned with the 
minor axis of the halo (which is generally in 
close alignment with the angular momentum axis).  
The first hypothesis is based recent attempts 
to reconcile the anisotropic of distribution 
of satellites in the Local Group 
\citep[e.g.][]{lynden-bell82,Hartwick00,
Willman_etal04,kroupa_etal05} with the 
predictions of numerical simulations 
\citep[e.g.][]{zentner_etal05,libeskind_etal05}.  
Alternatively, the simplest picture of disk 
galaxy formation predicts that halo and disk 
angular momenta should be nearly aligned 
\citep[][though cosmological hydrodynamic simulations 
show a misalignment of baryon and dark matter angular 
momenta that is typically $\sim 10^{\circ}-20^{\circ}$, 
see \citeauthor{vdb_etal02} \citeyear{vdb_etal02}; 
\citeauthor{chen_etal03} \citeyear{chen_etal03}]{fall_efstathiou80}.
We have defined the angle $\phi$ as the angle 
between the long axis of the primary disk and 
the position of the satellite in 
two-dimensional projection.  

\begin{figure}[t]
\epsscale{0.825}
\plotone{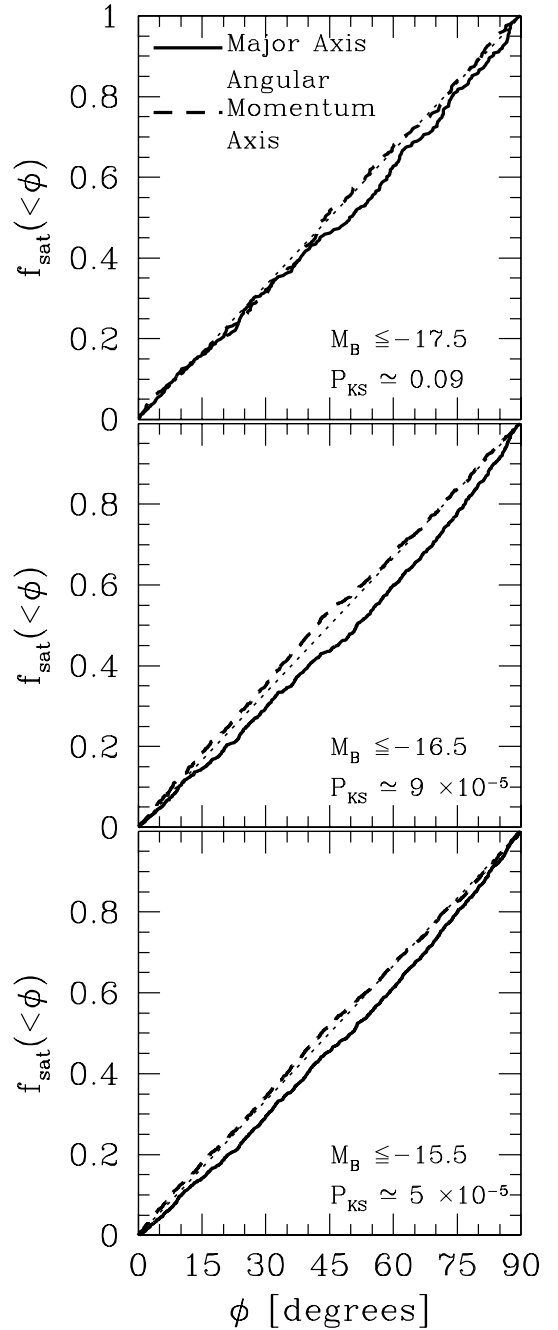}
\caption{
The cumulative angular distribution of 
satellites relative to the long axes of 
their primaries in simulations.  The 
three panels correspond to three different 
magnitude limits, $\bmag \le -17.5$, 
$\bmag \le -16.5$, and $\bmag \le -15.5$ from 
top to bottom.  
{\em Solid} lines correspond to the assumption 
that the disk rotation axis is aligned with 
the major axis of the primary halo.  
{\em Dashed} lines correspond to the assumption 
that the disk rotation axis is aligned with 
the angular momentum of the inner halo of the 
primary.  The {\em straight, dotted} lines 
correspond to an isotropic distribution in $\phi$.
In the lower right of each panel, we give the 
KS probability for the satellites in the 
samples with disk rotation along the major axis 
of the halo to be drawn from an 
isotropic distribution.  In all cases, the 
simulated samples with disk rotation aligned with 
halo angular momentum are consistent with isotropy.
}
\label{fig:fphi}
\end{figure}

The cumulative angular distributions are shown 
in Figure~\ref{fig:fphi}.  In the case 
where the disk is closely aligned with the 
major axis of the primary halo (solid lines), 
the anisotropy is apparent as a deficit 
of satellites at small and intermediate angles 
in Figure~\ref{fig:fphi}.  In this scenario, 
satellites are located preferentially near the 
short axes of their disk primaries in a sense similar 
to that observed by \citet{holmberg69}, 
\citet{zaritsky_etal97b}, and \citet{sales_lambas04}.  
A Kolmogorov-Smirnov (KS) test reveals that the 
probabilities for the simulated samples with 
$\bmag \le -16.5$ and $\bmag \le -15.5$ to be 
drawn from an isotropic underlying distribution are  
$\Pks \simeq~9 \times 10^{-5}$ and 
$\Pks \simeq~5 \times 10^{-5}$ respectively.  
In Figure~\ref{fig:fphi}, we also show the 
resulting distributions under the assumption 
that disk rotation is aligned with the 
angular momentum axis of 
the inner halo of the primary.  
Under this assumption, their appears to 
be a very slight excess of satellites at 
small and intermediate angles, but 
the distributions are consistent with 
isotropy.  The distributions with the 
disk angular momenta aligned with halo 
minor axes are quite similar to these 
and we have omitted them from 
Figure~\ref{fig:fphi} in the interest 
of clarity.

%
%

The next results concern the fraction of satellites 
that exhibit prograde or retrograde motion with respect 
to the disk of the primary galaxy.  In the case of the 
alignment of disk rotation with the major or minor 
axes of the host halo, there is no obvious way 
to assign a sense to the rotation to the disk.  
For this reason, we have computed prograde 
and retrograde statistics only for 
the case where the direction of the disk rotation 
is assumed to be aligned with the net 
angular momentum vector of the dark matter of 
the primary halo within $R < 0.3 \Rvir$.  
The results are summarized in 
Table~\ref{table:prsims}.  The error estimates 
listed in Table~\ref{table:prsims} represent the 
standard deviation of a binomial distribution 
with the appropriate parameters for each case.  
By presenting the prograde/retrograde statistics 
in this way, we have assumed that we can treat 
the three projections through the simulation as 
independent.  The excess of prograde satellites 
is a robust prediction in this scenario.  
Additionally, notice the trend for larger 
satellites to have a greater probability 
to exhibit prograde relative motion.  
For example, the satellites in the largest bin 
show a significant trend toward prograde motion 
as do the satellites with 
$-17.5 \ge \bmag \ge -18.0$, where there are 
$247$ satellites with $141$ progrades, yielding 
$\fprog = 0.57 \pm 0.03$.  Alternatively, in 
the magnitude range $-15.5 \ge \bmag \ge -16.0$, there 
are $294$ satellites, $152$ of which are 
prograde, yielding a prograde fraction in 
this bin of luminosity of $\fprog = 0.52 \pm 0.03$.  

%
%
%
%
%
%
\begin{table}[t]
\begin{center}
\caption{Prograde and Retrograde Satellites In Simulations 
\label{table:prsims}}
\begin{tabular}{llcccc}
\tableline\tableline
$\Vmax$ &
$\bmag$ &
$\Nsats$ &
$\Nprog$ &
$\Nretr$ &
$\fprog$ \\[0.5mm]
$\kms$ & & & & &
\\[2mm]
\tableline
$98.2$ & $-18.0$ & $81$ & $53$ &  $28$ & $0.65 \pm 0.05$ \\
$91.5$ & $-17.5$ & $328$ & $194$ & $134$ & $0.59 \pm 0.03$ \\
$86.4$ & $-17.0$ & $647$ & $360$ & $287$ & $0.56 \pm 0.02$ \\
$82.4$ & $-16.5$ & $925$ & $502$ & $423$ & $0.54 \pm 0.02$ \\
$78.8$ & $-16.0$ & $1226$ & $670$ & $556$ & $0.55 \pm 0.01$ \\
$75.9$ & $-15.5$ & $1520$ & $822$ & $698$ & $0.54 \pm 0.01$ \\
\\
\tableline
\end{tabular}
\end{center}
{\small
The prograde and retrograde fractions of satellites in the 
simulation volume.  Column description:  
(1) The minimum value of $\Vmax$ for satellites in the 
sample in $\kms$; (2) The corresponding maximum magnitude 
determined by matching halo number densities to 
galaxy number densities; (3) The total number of satellites 
in the sample, summed over the three, orthogonal projections 
through the simulation volume; 
(4) The number of prograde satellites 
that have peculiar velocities that are in 
the same direction as the rotation of the inner halo;  
(5) The number of retrograde satellites that have 
peculiar velocities that are in the opposite direction 
of the rotation of the inner dark matter halo; 
(6) The fraction of satellites with prograde velocities.
}
\end{table}
%
%

The excess of prograde satellites and the trend 
toward higher prograde fractions for larger satellites 
are not altogether surprising.  CDM halos tend to be connected 
through a network of filamentary structures established by 
the statistics of the density field and the 
tidal field during collapse in overdense regions 
\citep[e.g.][]{klypin_shandarin83,bond_myers96,
bond_etal96,colberg_etal05}.  
The matter distribution within the filaments is 
strongly concentrated toward the axis of the filament 
\citep{colberg_etal05} and halos preferentially 
tend to form in these dense regions and then merge along the 
dominant filamentary directions.  This leads to a strong 
correlation between directions along which halos 
merge during the mass accretion history of the primary halo 
\citep{knebe_etal04,zentner_etal05b}.  The subhalos that 
merge early and establish the angular momentum of the 
inner regions of the primary halo 
\citep[e.g.][]{vitvitska_etal02} flow along directions that 
are similar to the directions of infalling subhalos 
at late times, which are then observed as the satellites.  
Larger halos tend to be more strongly biased toward 
formation in the overdense regions 
\citep[e.g.][]{bond_myers96,bond_etal96}, 
so they are more faithful tracers of the 
flow along the filaments.  The higher prograde 
fractions for larger satellites are a 
reflection of this.

%
%
%
%
%
%
\begin{figure}[t]
\epsscale{1.0}
\plotone{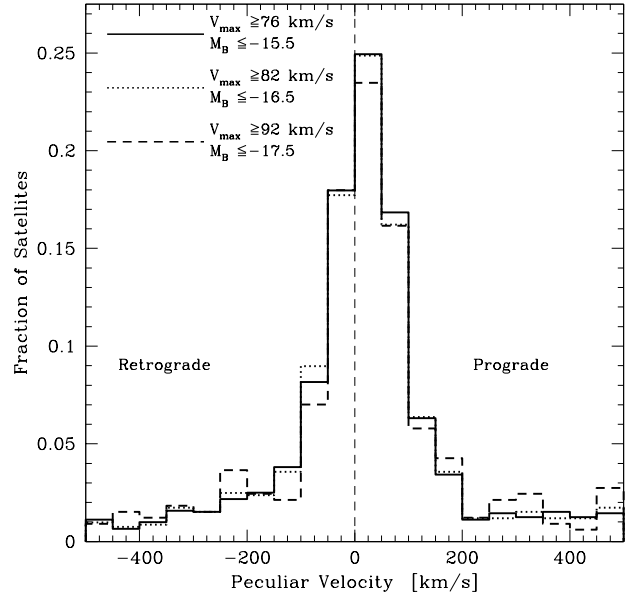}
\caption{
The peculiar velocities of the simulated satellite 
samples relative to the angular momenta of the 
central regions of the primary host halos.  
The histograms show the fraction of satellites in 
bins of peculiar velocity with retrograde 
satellites on the left (negative peculiar velocity 
relative to the disk) and prograde satellites on 
the right.  The {\em vertical, dashed} line shows 
the separation between prograde and retrograde 
satellites at zero peculiar velocity.  
We show histograms for three subsamples 
labeled by both their minimum $\Vmax$ and their 
maximum $\bmag$.  The {\em solid} line is 
$\Vmax \ge 76 \kms$ ($\bmag \le -15.5$), 
the {\em dotted} line is for 
$\Vmax \ge 82 \kms$ ($\bmag \le -16.5$), 
and the {\em dashed} line is for 
$\Vmax \ge 92 \kms$ ($\bmag \le -17.5$).
}
\label{fig:prhist}
\end{figure}
%
%

The excess of prograde satellites is also shown in 
Figure~\ref{fig:prhist}, which is a histogram 
of the peculiar velocities of the simulated satellite 
halos with respect to their primaries.  In this plot, 
positive peculiar velocities are defined to be 
in the same sense as the rotation of the disk.  
Again, the modest excess of prograde satellites, 
particularly with relatively small peculiar velocities, 
is apparent.

The simulated catalog provides a tool to 
study the effect of interlopers on the prograde/retrograde 
fractions because in this case we know both the ``observed'' 
sample and the true three-dimensional separation.  We have taken 
as a working definition to separate {\em true} satellites 
from {\em interlopers} a three-dimensional distance of 
$\rtrue = 1 \hMpc$.  Satellites that are a distance greater 
than $\rtrue$ from their primaries are considered interlopers.  
The virial radii of primaries in our sample range from roughly 
$175 \hkpc$ to $400 \hkpc$, so we generally expect objects 
at distances greater than $\sim 1 \hMpc$ to be unrelated 
to the primary.  In Table~\ref{table:proretnoi}, 
we show the prograde/retrograde fractions for the 
true satellites separated in three-dimensions by 
a distance less than $\rtrue$, with no contamination 
by interlopers.  We find that the net effect of interlopers 
at large line-of-sight separations is to dilute the 
prograde/retrograde fractions by $\sim 1-7\%$, 
depending upon the details of the sample selection.

%
%
%
%
\begin{table}[t]
\begin{center}
\caption{Prograde and Retrograde Satellites In Simulations with 
Interlopers Removed \label{table:proretnoi}}
\begin{tabular}{lcccc}
\tableline\tableline
$\bmag$ &
$\Nprog$ &
$\Nretr$ &
$\fprog$ &
$\ffalse$  \\[0.5mm]
\tableline
$-18.0$ & $ 42$ & $16$ & $0.72 \pm 0.06$ & $0.29$ \\
$-17.5$ & $145$ & $90$ & $0.62 \pm 0.03$ & $0.29$ \\
$-17.0$ & $278$ & $196$ & $0.59 \pm 0.02$ & $0.27$ \\
$-16.5$ & $385$ & $291$ & $0.57 \pm 0.02$ & $0.26$ \\
$-16.0$ & $499$ & $389$ & $0.56 \pm 0.02$ & $0.27$ \\
$-15.5$ & $608$ & $496$ & $0.55 \pm 0.02$ & $0.27$ \\
\\
\tableline
\end{tabular}
\end{center}
{\small
The prograde and retrograde fractions of satellites with 
true, three-dimensional distances smaller than 
$\rtrue = 1 \hMpc$ from their associated primaries 
in the simulation volume.  Column description:  
(1) The corresponding maximum magnitude 
determined by matching halo number densities to 
galaxy number densities; (2) The number of prograde satellites 
that have peculiar velocities that are in the same direction 
as the rotation of the inner halo;   
(3) The number of retrograde satellites that have 
peculiar velocities that are in the opposite direction 
of the rotation of the inner dark matter halo; 
(4) The fraction of prograde satellites; 
(5) The fraction of all satellites in 
Table~\ref{table:prsims} that were identified as 
``false'' satellites at each magnitude threshold, 
defined by the fact that they were included in 
the sample due to projection effects but 
are a distances greater than 
$\rtrue = 1 \hMpc$ from their primaries. 
}
\end{table}

%
%
\section{Observational Results}
\label{sec:observations}
\subsection{The Observed Sample}
\label{sub:sample}

Our observed sample of systems is extracted from ``Sample 2'' of 
\citet{prada_etal03}, which was constructed from SDSS data.  
Sample 2 has a maximum depth of $60000 \kms$ and a 
limiting absolute magnitude in $B$ of $\bmag = -18$.  
The isolation criteria of the primaries in the 
sample are that any satellites within a projected 
radial separation of $\rp \le 500 \hkpc$ and with a 
line-of-sight velocity difference 
$\Delta V \le 1000 \kms$ must be dimmer than the 
primary by at least $\Delta \bmag \geq~2$.  
Satellites must satisfy $\Delta \bmag \geq~2$ with respect
to their associated primary, they must have a projected
distance to the primary of 
$\rp \le 350 \hkpc$, and a line-of-sight velocity 
difference of $\Delta V \le 500 \kms$.  

Sample 2 contains more than $700$ primaries and more 
than $1000$ satellites.  Using Sample 2 as a starting point, 
we selected our primaries according to the following additional 
criteria.
\begin{enumerate}

\item We enforced a limiting depth of $11000 \kms$.

\item We chose primaries only in the magnitude range 
$-20.5 \le \bmag \le -19.5$

\item We required all primaries to have a 
well-defined morphological type as a spiral.

\end{enumerate}
The limit of $11000 \kms$ was imposed not to lose too 
many satellites, as the SDSS is complete down to 
$\bmag = -17.5$ at $11000 \kms$, thus the satellites
(two magnitudes fainter than primaries) are those 
brighter than $\bmag = -15.5$ only.  
The narrow magnitude range for the primaries was chosen 
to avoid including primaries with very different masses,
given the known dependence of luminosity on the mass
of the primary \citep[e.g.][]{prada_etal03}.  
Further, satellites 
with projected distances from their primaries 
smaller than $20 \kpc$ were rejected,
due to potential galaxy misclassification 
(e.g. HII region contamination, bright stars etc.).  
After applying these constraints, we were 
left with $144$ spiral primaries and $193$ satellites.
We obtained the direction of rotation of 
a limited number of primaries, so we analyzed the prograde 
and retrograde motion using a subsample of $43$ primary 
galaxies with $76$ satellites (see \S~\ref{sub:obs_spin}).
For the angular distribution analysis, we used the
entire sample of $193$ satellites (see \S~\ref{sub:ang_dist}). 

%
%
\begin{table*}[t]
\begin{center}
\caption{The Primaries and Satellites of the sample. \label{table:obstable}}
\begin{tabular}{ccccccccc}
\tableline
Name       \tablenotemark{1} &
Bmag       \tablenotemark{2} &
RA(2000)   \tablenotemark{3} &
DEC(2000)  \tablenotemark{3} &
PA         \tablenotemark{4} &
V($\kms$)  \tablenotemark{5} &
$\Delta V$ \tablenotemark{6} &
dist (kpc) \tablenotemark{7} &
Orbit      \tablenotemark{8} \\
\tableline
     pgc001112 & -20.3 & 00 16 55 & -00 05 18 & 170 & 10042.9 &      -   &     -   &   - \\
   {\it a}     & -18.3 & 00 17 07 & -00 08 30 & 137 & 10098.7 &     55.8 &   175.1 &   r \\
     pgc001841 & -20.1 & 00 30 07 & -11 06 49 & 169 &  3649.6 &      -   &     -   &   - \\
   {\it a}     & -15.9 & 00 31 28 & -10 40 33 &  37 &  3645.2 &     -4.4 &   486.0 &   p \\
  mcg-2-3-61   & -20.2 & 01 00 04 & -11 04 56 & 123 &  5519.4 &      -   &     -   &   - \\
   {\it a}     & -17.9 & 01 00 05 & -11 02 32 &   6 &  5420.0 &    -99.4 &    52.2 &   r \\
    ngc341     & -20.4 & 01 00 45 & -09 11 08 &  55 &  4658.0 &      -   &     -   &   - \\
   {\it a}     & -16.9 & 01 01 43 & -09 19 00 & 119 &  4639.3 &    -18.7 &   294.5 &   p \\
    ugc1962    & -20.0 & 02 28 54 &  00 22 13 & 105 &  6152.9 &      -   &     -   &   - \\
   {\it b}     & -17.4 & 02 29 33 &  00 22 23 &  91 &  6549.3 &    396.4 &   240.4 &   p \\
   {\it c}     & -17.0 & 02 29 39 &  00 07 23 &  37 &  6155.4 &      2.5 &   455.2 &   p \\
\end{tabular}
\tablenotetext{1}{Host names and satellite labels}
\tablenotetext{2}{Absolute magnitude in B}
\tablenotetext{3}{J2000 coordinates of the objects}
\tablenotetext{4}{Sky PA of the hosts and, for the satellites,
                  the Position Angle of their location
                  with respect to the host}
\tablenotetext{5}{Recessional velocity ($h = 0.7$)}
\tablenotetext{6}{The difference in recessional velocity}
\tablenotetext{7}{Projected distance between satellite and host}
\tablenotetext{8}{``p'' stands for prograde and ``r'' for retrograde.}
\tablecomments{This table is published in
its entirety in the electronic edition of the Astrophysical Journal.\\
A portion is shown here for guidance regarding its form and content.}
\end{center}
\end{table*}
%
%

The observational data listed in Table~\ref{table:obstable}
come from the SDSS database and from our own observations with 
the exception of NGC2841, which had a previously-published 
rotation curve that we took from \citet{afanasiev_silchenko99}.  
Data such as recessional velocities, 
absolute magnitudes and positions for primaries and 
satellites were taken from the SDSS database, while the
Position Angles of the primaries were taken from the 
Lyon-Meudon Extragalactic Database (LEDA)
\footnote{{\tt URL http://leda.univ-lyon1.fr}}.  
Low-resolution spectroscopy of the 
H$\alpha$ line easily determines the spin direction 
of our objects.  Therefore, through an observational 
program at the telescopes of the Isaac Newton Group 
(La Palma, Canary Islands), we obtained rotational
data for $42$ of our primaries.

Observations were performed using the spectrographs ISIS, 
of the $4.2$~m William Herschel Telescope (WHT), 
and IDS, of the $2.5$~m Isaac Newton Telescope (INT) 
of the Isaac Newton Group, on La Palma, Spain.  
The setup was simple and no special 
observing conditions were needed, 
so most of the data were taken in service mode.  
The most important aspect of the observation 
was to determine reliably the orientation of the
charge-coupled device (CCD), to be able to place 
the receding end of the galaxies correctly on the sky.  
Two exposures at Sky PA $0^{\circ}$, shifted 
by some $10\arcsec$, and two more 
at Sky PA $90^{\circ}$ with the same
offset were sufficient to mark 
North and East on the images.

We used the R316R or R600R gratings on the Marconi II chip 
with ISIS, centered at $6650 \AA$, 
and R900V, R600V on chip EEV13 or R1200R on chip
EEV10 with IDS, also centered at $6650 \AA$.  
The slit width ranged from $1\arcsec$ to $1.5\arcsec$, 
depending on the seeing conditions.  
Exposure times were typically $300 \seconds$ with WHT 
and $900 \seconds$ with INT.  During the observations, 
the seeing varied from $0.8\arcsec$ up to $2.5\arcsec$.  
The detection of the shift on the 
H$\alpha$ line was performed by analyzing 
the images with the IRAF package, 
comparing the line centroid shift 
with that of a sky line to correct 
for spurious rotations of the CCD.
The observing runs took place over the 
period from March to September of 2003.
%
%
%
%
%
%
%
\subsection{The Angular Distribution of Satellites}
\label{sub:ang_dist}

In order to study the distribution of the angular positions 
of the observed satellites relative to their 
disk primaries we were able to use our largest sample of 
objects because this requires only the coordinates 
of all objects and the Position Angles of the primaries, 
which we extracted from the LEDA database.  
We included in this analysis 
the $144$ disk primaries in the magnitude bin 
$-20.5 \le \bmag \le -19.5$, and the $193$ satellites 
selected according to the description in \S~\ref{sub:sample}.  
We reduced the angular distance of the satellites
from the disks to a single quadrant, 
$\phi \in [ 0^{\circ}, 90^{\circ} ]$. 

%
%
\begin{figure}[t]
\epsscale{1.0}
\plotone{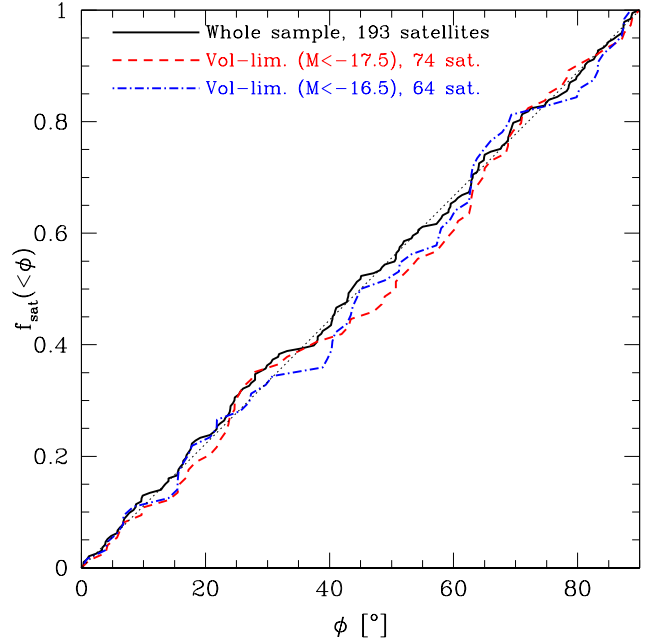}
\caption{
The cumulative distribution of the angular positions
of satellites relative to their primary disk galaxies.
The samples shown in this plot are the entire sample with 
$193$ satellites and $144$ primaries (solid line), 
a volume-limited subsample at a depth of $11000 \kms$ 
with $74$ satellites and $62$ primaries (dashed line)
and a volume-limited subsample at a depth of $7000 \kms$ with $64$
satellites and $51$ primaries (dot-dashed line).  The 
{\em thin, dotted} line in the background represents 
an isotropic cumulative distribution function.
}
\label{fig:angdist}
\end{figure}

Figure~\ref{fig:angdist} shows the cumulative 
frequency for the entire satellite sample (solid line), 
compared to that of an isotropic distribution (dotted line).  
In addition, we computed the angular distributions of 
satellites in two volume-limited subsamples.  
The first subsample has a depth of $11000 \kms$ 
and limiting magnitude $\bmag = -17.5$, and 
the second has a depth of $7000 \kms$ and limiting 
magnitude $\bmag = -16.5$.  
The statistics for the volume-limited subsamples
are poorer ($74$ satellites at $11000 \kms$ and 
$64$ at $7000 \kms$), but the behavior is qualitatively 
similar to that of the entire sample of $193$ satellites.

At first glance, this result seems most easily compatible 
with the simulated distribution with disk rotation nearly 
aligned with the angular momentum of the 
inner halo (Figure~\ref{fig:fphi}).  We have performed a 
comparison of the distributions, using a KS test to 
test several hypotheses.  
The full observational sample as well as the volume-limited 
subsamples are all consistent with an isotropic 
distribution about their primaries.  Contrary to 
the results of \citet{zaritsky_etal97b} and 
\citet{sales_lambas04}, this sample {\em does not} 
show significant evidence of anisotropy.  In addition, 
the observed sample is consisted with all three of 
the hypotheses used to construct 
distributions from the simulation.  This includes the 
case with the disk rotation axis aligned with the 
major axis of the host halo, 
which yields a probability $\Pks \simeq~0.17$ of being 
drawn from the same underlying distribution.  The 
relatively small size of the current observational sample 
makes it difficult to reject any of our idealized 
disk alignment hypotheses.  

%
%
%
%
%
\subsection{Observed Prograde Satellite Fractions}
\label{sub:obs_spin}

To state if a satellite is prograde or not, 
we need to know the difference in recessional 
velocity between the satellite and the primary
($\Delta v = V_{sat} - V_{prim}$), the
Position Angle of the satellite with respect to the 
primary ($PAS$), and  the sky Position Angle  
of the redshifted end of the primary disk ($PAred$). We
define the quantity $S$ by 
\begin{eqnarray}
\label{eq:S}
S = +1 & \mathrm{ if } & 
\left\{ \begin{array}{l}
\mid PAred - PAS \mid < 90^{\circ} \\
\mid PAred - PAS \mid > 270^{\circ}\end{array}\right\} \nonumber\\
S = -1 & & \mathrm{ otherwise, } \nonumber
\end{eqnarray}
A satellite is prograde if $S \cdot \Delta v > 0$,
and retrograde $S \cdot \Delta v < 0$.  
Table~\ref{table:obstable} gives the sense of the 
line-of-sight velocities of the satellites in 
our sample with respect to their disk primaries.

\begin{table}[t]
\begin{center}
\caption{Prograde/Retrograde Statistics of the Entire Observational Sample 
\label{table:proret11000}}
\begin{tabular}{ccccc}
\tableline\tableline
Angle from disk & $\Nsats$  & $\Nprog$ & $\Nretr$ & $\fprog$ \\
\tableline
$0^{\circ}$ to $+30^{\circ}$ & 24 & 15 & 9 & $0.63 \pm 0.10$ \\
$0^{\circ}$ to $+45^{\circ}$ & 40 & 25 & 15 & $0.63 \pm 0.08$ \\
$0^{\circ}$ to $+75^{\circ}$ & 58 & 38 & 20 & $0.66 \pm 0.06$ \\
$0^{\circ}$ to $+90^{\circ}$ & 76 & 46 & 30 & $0.61 \pm 0.05$ \\
\\
\tableline
\end{tabular}
\end{center}
{\small  
Statistics of prograde/retrograde 
satellites for the entire sample of satellites.
The columns are (1) the angular displacement
from the primary disk plane, summed over the
four quadrants; (2) the number of satellites
for each angular selection; (3) the number of
prograde satellites found for each angular
selection; (4) the number of retrograde
satellites found for each angular selection;
(5) the frequency of prograde satellites
over the total with an error estimate given 
by the standard deviation of the appropriate 
binomial distribution.
}
\end{table}
%
%

%
%
%
\begin{table}[t]
\begin{center}
\caption{ Prograde/Retrograde Statistics for the 
Entire Sample to a Depth of $V < 7000 \kms$ (upper panel) and
for a Volume-limited Sample to a Depth of $V < 7000 \kms$
(lower panel) \label{table:proret7000} }
\begin{tabular}{ccccc}
\tableline\tableline
Angle from disk & $\Nsats$ & $\Nprog$ & $\Nretr$ & $\fprog$\\
\tableline
$0^{\circ}$ to $+30^{\circ}$ & 11 &  8 & 3 & $0.72 \pm 0.13$ \\
$0^{\circ}$ to $+45^{\circ}$ & 20 & 14 & 6 & $0.70 \pm 0.10$ \\
$0^{\circ}$ to $+75^{\circ}$ & 34 & 22 & 12 & $0.64 \pm 0.08$ \\
$0^{\circ}$ to $+90^{\circ}$ & 41 & 26 & 15 & $0.63 \pm 0.08$ \\
\tableline\tableline
Angle from disk & $\Nsats$ & $\Nprog$ & $\Nretr$ & $\fprog$\\
\tableline
$0^{\circ}$ to $+30^{\circ}$ &  9 &  6 & 3 & $0.66 \pm 0.16$ \\
$0^{\circ}$ to $+45^{\circ}$ & 13 &  8 & 5 & $0.61 \pm 0.13$ \\
$0^{\circ}$ to $+75^{\circ}$ & 23 & 13 & 10 & $0.57 \pm 0.10$ \\
$0^{\circ}$ to $+90^{\circ}$ & 28 & 16 & 12 & $0.57 \pm 0.09 $ \\
\tableline
\end{tabular}
\end{center}
{\small Prograde/Retrograde statistics of 
satellite galaxies for an unlimited subsample at a 
depth of $7000 \kms$ and for a corresponding volume-limited 
subsample at the same depth.  
The depth of $7000 \kms$ ($\bmag = -16.5$) has been 
chosen in an attempt to keep a significant 
number of object in each subsample.
The columns are (1) the angular displacement
from the primary disk plane, summed over the
four quadrants; (2) the number of satellites
for each angular selection; (3) the number of
prograde satellites found for each angular
selection; (4) the number of retrograde
satellites found for each angular selection;
(5) the frequency of prograde satellites
over the total.
}
\end{table}

The resulting prograde satellite fractions are 
summarized in Table~\ref{table:proret11000} and 
Table~\ref{table:proret7000}.  Unfortunately, the 
small size of the observational sample does not allow 
us to study any magnitude dependence of the prograde 
satellite fraction at a statistically-significant 
level as we did for the simulated 
satellites in \S~\ref{sub:simresults}.  
Alternatively, we present the prograde/retrograde 
fractions as a function of angular displacement from 
the major axis of the primary galaxy disk.  
Table~\ref{table:proret11000} shows the 
resulting distribution of prograde and retrograde 
satellites for the entire sample at full depth 
($11000 \kms$).  
Table~\ref{table:proret7000} shows two subsamples 
with limiting depths of $7000 \kms$, one containing 
all objects to this depth and one for a 
volume-limited subsample with an absolute magnitude limit 
of $\bmag = -16.5$.

It is evident from Table~\ref{table:proret11000} that 
the largest samples of satellites exhibit an excess 
of prograde peculiar velocities that is significant 
at the $\sim 2-2.5\sigma$ level.  In addition, the 
data show no evidence for any significant trend in 
the prograde fraction as a function of the angular 
position with respect to the disk.  At all angles 
with respect to the disk, the prograde fraction is 
$\sim 60\%$, though the statistical significance of 
the prograde excess in the smallest angular cuts is marginal.  
Table~\ref{table:proret7000} shows a qualitatively 
similar result, with $\fprog \sim 60\%$ in both of these 
subsamples, though again, the statistical 
significance is, at best, marginal in these small samples.  
In addition to these results we also found no significant 
evidence for a variation in the prograde fraction as a 
function of projected distance to the primary.  

\begin{figure}[t]
\epsscale{1.0}
\plotone{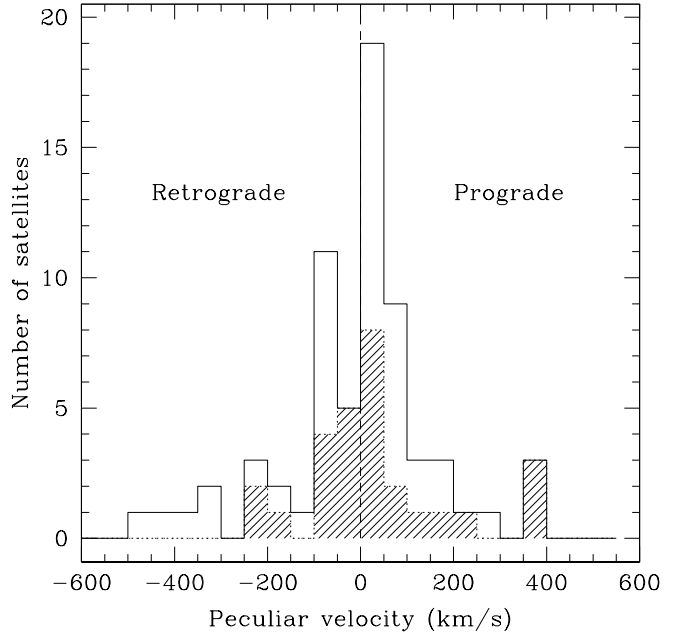}
\caption{
The distribution of peculiar velocities of the 
entire unlimited sample (solid line, empty), and the
volume-limited sample at a depth of $7000 \kms$ (dotted line, dashed).
Positive peculiar velocities correspond to 
prograde satellites and negative peculiar 
velocities correspond to retrograde satellites.
The zero peculiar velocity is marked by a vertical
dashed line.
}
\label{fig:histproret}
\end{figure}

Figure~\ref{fig:histproret} shows a histogram 
of the peculiar velocities of two of our samples, the
entire sample with all objects and the volume-limited
subsample at a depth of $7000 \kms$ ($\bmag \le -16.5$).  
Figure~\ref{fig:histproret} shows a peak in the 
slowly-rotating progrades in both of these samples.  
This seems to be in broad agreement with the results of 
the simulated sample of subhalos shown in 
Figure~\ref{fig:prhist}.  A more detailed analysis would 
require a larger observational sample.

%
%
%
%
%
%

%
%
%
%
%
%

\section{The Effect of Observational Biases}
\label{sec:biases}

We performed a series of Monte Carlo simulations of mock
observational samples in order to quantify the 
effects of observational biases and interlopers.
We produced lists of objects, both satellites and interlopers,
distributed in magnitude according to the 
luminosity function of each class of object.  
This analysis complements the previous results on 
the net effect of interloper contamination in the 
cosmological $N$-body simulation performed in 
\S~\ref{sub:simresults}), because it allows us to 
study a statistically-large number of independent, 
mock observations for a variety of assumptions about the 
true fraction of prograde satellites.

The first set of Monte Carlo simulations we performed 
were aimed at assessing the effect of interlopers (which 
are separated from their primaries by large distances 
along the lines of sight, but are associated due to projection) 
on the measured prograde/retrograde fractions of satellites 
relative to the prograde/retrograde fractions of ``true'' 
satellite galaxies that have small three-dimensional 
separations from their primaries.  
The number of objects at each bin 
of magnitude was determined using 
\citet{schechter76} luminosity functions with 
different parameters according to each class of 
object.  Specifically, we employed 
the magnitude form of the Schechter function, 

%
\begin{equation}
\label{eq:schm}
\Phi(M) =  0.4\Phi_{*}
\log(10)\ 10^{0.4(M_{*} - M)(\alpha + 1)}
\exp({-10^{0.4(M_{*} - M)}}),
\end{equation}
which we then used to determine the number 
of objects by integrating over magnitude and volume.

For the interlopers, we used the parameters given by 
\citet{blanton_etal03} for SDSS field galaxies, converted to
B band (see section~\ref{sub:methods}), namely 
$\Phi_{*} = 0.0074 \Mpc^{-3}$, $M_{*} = -21.31$, $\alpha = -0.89$.  
In this case, the volume we considered was the same truncated 
cone as in our real observations, but with an additional constraint 
that the distance $r$ from the primary satisfies $r > 1\hMpc$.  
This constraint serves as our working definition of ``interloper'' as 
opposed to ``true'' satellite.

For the satellites, we used the parameters of the 
Local Group (LG), assuming that it represents 
a typical system of satellite galaxies 
\citep{mateo98,vandenbergh00}.  
The Schechter function parameters for the LG
satellites were $M_{*} = -19.5$ and  $\alpha = -1.14$.
However, the characteristic number density $\Phi_{*}^{sat}$
for the LG is not well defined as it depends 
sensitively on the volume considered.  
A way to get a first approximation of its value
is to assume a given interloper contamination, which
can be taken from our Table~\ref{table:prsims} in
\S~\ref{sub:simresults}, and is in agreement with 
the fraction of interlopers found by \citet{prada_etal03} 
from their satellite dynamics analysis.
We then assume an interloper contamination of $25\%$
and calculate $\Phi_{*}^{sat} $ such that we recover
$25\%$ of interloper contamination in the simulation. We used
$\Phi_{*}^{sat} \simeq~0.07 \Mpc^{-3}$. The volume
considered for the satellites was the intersection of 
the truncated cone of our real observations with the 
sphere contained within $r < 1\hMpc$ about each primary.

We then observed the objects in our simulation in the 
same way as we did for the SDSS data. For simplicity, 
in our first suite of Monte Carlo mock observations, 
we kept a fixed recessional velocity of $7000 \kms$ for all the 
objects in these mock catalogs.
We assumed that interlopers always have a $50\%$ 
chance of being prograde or retrograde, while
the prograde fraction of satellites and the total number 
of primaries are both input parameters that are used 
to construct specific mock observations.
The number of satellites for each galactic system (or
for each primary) is then calculated using the
Schechter function with the $\Phi_{*}^{sat}$ as above.

We run $500$ Monte Carlo trials, with an input 
prograde fraction of $\fprog = 60\%$;
the input number of primary galaxies was selected
such that the total number of combined false and true satellites
in the mock sample was about $76$ as in our real observations.
The interloper contamination in our mock samples reduced the
observed prograde fraction to an average of $\fprog = 0.57$ 
with a root-mean-square scatter about the mean of $\pm 0.05$.
This agrees quite well with the effect of interlopers 
found in the mock catalogs constructed from the halos in 
the cosmological $N$-body simulation 
(see Tables~\ref{table:prsims} and \ref{table:proretnoi}).

The second set of Monte Carlo simulations was aimed at 
determining how the prograde/retrograde fractions are affected by 
having systems at different recessional velocities mixed together,
as we have in our observed sample.  When several galactic
systems at different recessional velocities are mixed together, 
different levels of interloper contamination are superposed.
Also, the apparent magnitude limit of the observations results in 
a selection of satellite luminosities that depends on the 
recessional velocities of the systems.  The aim is 
to simulate these two observational effects.
We used the same setup as for the Monte Carlo simulations 
with fixed recessional velocities that we discussed 
in the previous paragraph.  
We built up three samples of systems, again with the necessary
input number of primaries so as to ``observe'' a total of $76$
false and true satellites combined, at recessional velocities of
$2800, 5600$ and $8800 \kms$.  We then observed all of the
samples together as a single, mock sample.

\begin{figure}[t]
\epsscale{1.0}
\plotone{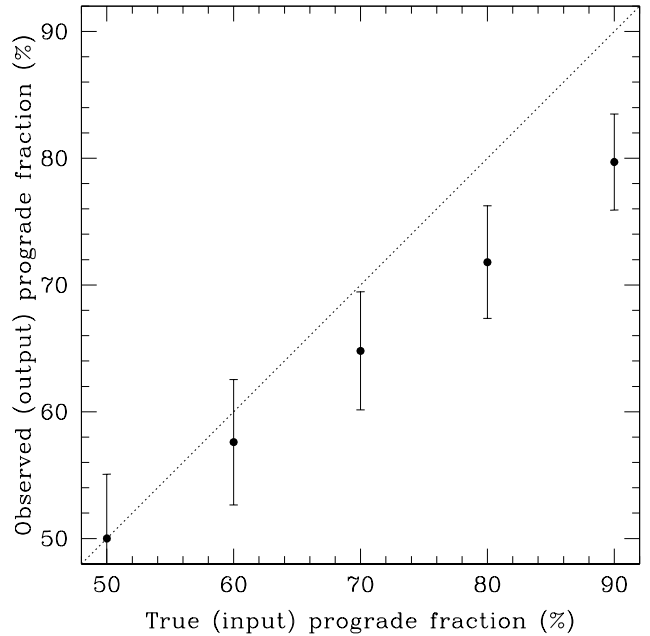}
\caption{
Observed reduction of the prograde fraction caused by 
interloper contamination in a sample containing systems 
at a wide range of recessional velocities.  
The horizontal axis represents the true prograde fraction for all 
true satellites and the vertical axis represents the observed 
prograde fraction taking interloper contamination into account.
The {\em points} represent the mean observed prograde fraction 
derived from our Monte Carlo simulations of the net 
effect of interloper contamination in the case of mixing systems
with different recessional velocities. The error
bars represent the root-mean-square scatter about the mean.  
The {\em dotted} line represents perfect retrieval of the prograde 
fraction.
}
\label{fig:low}
\end{figure}

We ran a series of $500$ such random trials with an 
input satellite prograde fraction of $\fprog = 60\%$ 
as above.  
The mean observed percentage of prograde satellites 
in the mock samples was $\fprog \simeq 58\%$, 
with a root-mean-square scatter of 
$\simeq 5\%$ about the mean.  
This implies that the systematic dilution 
of the prograde/retrograde fraction due to interlopers is 
fairly small for comparable samples; however, the scatter 
about the true fractions due to the presence of interlopers 
is not insignificant for samples of this size.  
We show in Figure~\ref{fig:low} the mean reduction of the
``observed'' prograde fraction as a function of the true
input prograde fraction as well as the $1\sigma$ scatter 
in the observed values about the mean.
%
%
%
%
%
\vspace*{1cm}
%
%
%
%
\section{Conclusions}
\label{sec:results}

We study the distribution of satellite galaxies 
about disk primaries.  The study has both observational and 
theoretical components.  We focus our attention on the 
fraction of satellite galaxies that exhibit prograde motion with
respect to their disk primaries and on the angular distribution 
of satellites relative to their primaries.  On the observational side,
we study a carefully-selected observational sample of primary 
disk galaxies and satellite galaxies extracted from the 
Sloan Digital Sky Survey database.  
In addition, we analyze the structure and distribution of halos
and subhalos in an $N$-body simulation of structure formation in the, 
now standard, concordance $\Lambda$CDM model.

We find that the fraction of prograde satellites in our entire 
flux-limited sample is $\fprog = 0.61 \pm 0.05$. 
The observational samples show no evidence for a strong 
dependence of $\fprog$ on magnitude, angular distance 
from the satellite to the major axis of the primary disk, 
or projected distance from the primary.  However, the 
observed samples are, as yet, small and as a result, 
the statistics are poor.  In addition, we found that 
the observed satellite galaxies are consistent with 
being distributed isotropically about their primary disk 
galaxies.

From the cosmological $N$-body simulation, we
constructed catalogs of primaries and satellites 
using an algorithm that maps galaxies onto halos 
and subhalos based on matching number density 
\citep[e.g.][]{kravtsov_etal04}.  
We built catalogs based on three idealized hypotheses 
for the orientations of disk galaxies within
their surrounding dark matter halos, with the
angular momentum of the disk aligned with either (1) the major axis
of the inner host halo, (2) the minor axis of the host halo, 
or (3) the angular momentum of the particles in the inner host halo.  
We then presented the general predictions of these scenarios.  
The first scenario predicts a projected distribution of 
satellite galaxies that is strongly anisotropic, with
galaxies located near the poles of their disk primaries.  For 
example, the Kolmogorov-Smirnov probability that the entire 
simulated distribution of halos that are assigned magnitudes 
$\bmag \le -15.5$ is selected from an underlying isotropic 
distribution is $\Pks \simeq~5 \times 10^{-5}$.  
However, a KS test does not yet allow us to reject this
hypothesis based on the present data, largely because the 
observational data set is fairly small 
($193$ satellites in our largest, flux-limited sample).  
The other two hypothesized disk orientations predict 
satellite distributions that are consistent with isotropy 
at all magnitudes.  Under the hypothesis that the angular 
momenta of the disk primaries are aligned with the angular 
momenta of the inner regions of the dark matter halos that 
host them, the simulation predicts a prograde satellite fraction 
$\approx~55-60\%$.  Specifically, for the entire sample of 
satellites with $\bmag \le -15.5$, we find 
$\fprog = 0.54 \pm 0.01$, while for a subsample of the 
largest satellites with $\bmag \le -17.5$, we find 
$\fprog = 0.59 \pm 0.03$.  As the previous sentence 
indicates, we find that $\fprog$ is a weak function
of satellite size (magnitude), whereby larger (more luminous) 
satellites are more likely to exhibit prograde motions.  
The simulation results are broadly consistent with the values
measured in the observational sample.  
A larger observational sample will be needed in 
order to test for any magnitude dependence in the 
observed prograde satellite fraction and to make 
more meaningful comparisons between theoretical 
predictions and observations.  Larger observational 
samples will also require a significant refinement 
of theoretical predictions.

In addition to these results, we also made an effort to assess 
the net effect of interlopers on the observed prograde satellite 
fractions compared to the prograde satellite fractions for ``true'' 
satellites.  From our analysis of both the $N$-body simulation 
and an independent set of Monte Carlo simulations of mock observations,
we found that the mean dilution of the prograde satellite fraction 
due to interloper contamination is of the order of a few percent.  
However, we found that the scatter in the observed prograde and 
retrograde fractions in samples comparable to the size of our 
observational set is sizable at nearly $5\%$.  

A comparable observational study including the 
prograde/retrograde satellite fractions 
and the angular distributions of satellites 
was performed by \citet{zaritsky_etal97b,zaritsky_etal97} who 
had a slightly larger sample of systems.  Contrary to our findings, 
\citet{zaritsky_etal97} reported more retrograde satellites than 
prograde satellites ($f_{\mathrm{prog}} \simeq 48\%$).  However, 
note that this result is not statistically significant as 
an estimate of the statistical error on the mean from a 
sample of the size of \citet{zaritsky_etal97b} 
gives $\sigma(\fprog) \simeq 5\%$.  
\citet{zaritsky_etal97b} also reported strong evidence for 
an anisotropic satellite distribution as we mentioned in 
\S~\ref{sec:intro}.  We were unable to confirm this result 
with our observed satellite sample, which is consistent with 
an isotropic underlying distribution.  
The nature of any disagreements between our results and 
the previous work of \citet{zaritsky_etal97b,zaritsky_etal97} 
is unclear at present; however, 
it could possibly be due to different 
criteria in the selection of the objects, unaccounted for some
observational biases, and/or statistical fluctuations in the 
relatively small samples.  Larger samples constructed from 
forthcoming data sets should be able to address these issues 
as well.

\acknowledgments

We are grateful to Brandon Allgood for help with the simulation 
analysis and helpful discussions.  We thank Juan Betancort, 
Carlos Guti\'errez, Andrey Kravtsov, and Risa Wechsler for valuable discussions
throughout the course of this work.  We thank the Isaac Newton Group 
personnel at La Palma (Canary Islands) for their support during 
the course of our observational program.  ARZ is supported by the Kavli 
Institute for Cosmological Physics at 
The University of Chicago and the National Science Foundation under 
grant NSF PHY 0114422.  AAK is supported by the NSF 
grant AST-0206216 to New Mexico State University.  
The simulations analyzed in this work were performed on 
Columbia at the National Aeronautics and Space Administration (NASA) 
Advanced Supercomputing Center.  We acknowledge use of the 
Lyon-Meudon Extragalactic Database (LEDA).  
This work made use of the NASA Astrophysics Data System.


\vspace*{2cm}
\bibliography{ms}
\end{document}